\documentclass[preprint2]{aastex62}
\usepackage[utf8]{inputenc}
\usepackage{graphicx}
\usepackage{mathtools}

\begin{document}
\title{Searching for Systematics in Forward Modeling Sunyaev-Zeldovich Profiles}
\author[0000-0003-1593-1505]{Emily Moser}
\affiliation{Department of Astronomy, Cornell University, Ithaca, NY 14853, USA}

\author[0000-0001-5846-0411]{Nicholas Battaglia}
\affiliation{Department of Astronomy, Cornell University, Ithaca, NY 14853, USA}

\author[0000-0002-4200-9965]{Stefania Amodeo}
\affiliation{Université de Strasbourg, CNRS, Observatoire astronomique de Strasbourg, UMR 7550, F-67000 Strasbourg, France}

\begin{abstract}
    The Sunyaev-Zeldovich thermal (tSZ) and kinetic (kSZ) effects can be used to constrain the thermodynamic properties of pressure and density, respectively, of galaxies and their surrounding regions. As SZ observations continue to improve, it is important to understand any modeling systematics when inferring properties from the data. Thus, a pipeline to forward model observed SZ profiles was developed called \texttt{Mop-c-GT}. Previous studies have used this repository to create modeled SZ profiles by selecting halos from the IllustrisTNG simulation and found significant differences between the simulated profiles and those recently measured by the Atacama Cosmology Telescope. There are many uncertainties involved in modeling observed samples and in the forward modeling process, so in this study, we explore methods implemented in \texttt{Mop-c-GT} and in the selection of the simulated halos to test the effects on the resulting modeled profiles. After testing several methods within the forward modeling process and varying the halo selection from the simulation, we find minimal differences between the simulated tSZ profiles of the original calculation and the updated methods, indicating that the observations still pose a challenge for the numerical methods used to describe the astrophysics of these systems.
\end{abstract}

\section{Introduction}
The regions surrounding galaxies, known as the circumgalactic medium (CGM) and intergalactic medium (IGM), play integral roles in the change of the galaxy over time. The CGM and IGM supply the galaxy with baryonic material for astrophysical processes and act as reservoirs into which feedback mechanisms within the galaxy (e.g., supernovae and active galactic nuclei) expel material to be recycled. A widely-studied problem in extragalactic astronomy is to further understand the thermodynamic processes occurring within the CGM and/or IGM and the interactions with the host galaxy to shed light on galaxy evolution. 

An emerging probe of the CGM is the Sunyaev-Zeldovich (SZ) effect \citep{SZ1970} of the cosmic microwave background (CMB). As CMB photons propagate through these halos they are inversely Compton-scattered off ionized electrons, leading to an increase in high-energy photons and decrease of low-energy photons. This is known as the thermal (tSZ) effect \citep{SZ1972}, and since it is proportional to the line-of-sight (LOS) integral of the electron pressure, it can be used as a probe of the gas pressure within the halos. Additionally, the kinetic (kSZ) effect \citep{SZ1980} describes the Doppler shift of CMB photons scattering off free electrons in galaxies and galaxy clusters with peculiar velocities. This effect leads to shifts in the CMB temperature that are proportional to the LOS integral of the peculiar velocity multiplied by the electron number density, and can therefore be used as a probe of the gas density within the halos. Combining the density and pressure information from these SZ effects provides complete thermodynamic information of ionized gas in the CGM \citep{Battaglia2017} and can thus provide constraints on the processes governing galaxy evolution.

We have had the technology to observe the tSZ effect for several years, both in cluster and galaxy mass scales (e.g., \citet{Hasselfield2013,Bleem2015,Greco2015,Planck_tsz2016,Huang2020,Hilton2021,Meinke2021,Reichardt2021,Vavagiakis2021,Bleem2022,Bregman2022}) The kSZ effect on the other hand has only recently been detected in the last decade, first through cross-correlations of CMB observations with galaxy surveys \citep{Hand2012} and since then, with many CMB experiments and galaxy catalogs along with individual galaxy clusters (e.g., \citet{Sayers2013,Planck_kSZ2016,Schaan2016,Soergel2016,Adam2017,Tanimura2021,Reichardt2021,Calafut2021,Chen2022}). 

With the improvement in the signal-to-noise of tSZ and kSZ observations over the past decade, we are now able to combine these observations to provide constraints on the thermodynamic processes of these systems (e.g., \citet{Schaan2021,Amodeo2021}). However, as instrument sensitivity continues to improve, it is becoming increasingly important to understand any underlying systematics in interpreting the data. Therefore, it is necessary to develop methods to forward model tSZ and kSZ profiles most closely matching known properties of profiles observed by a given experiment. This includes both appropriately modeling properties of the observed sample, as \citet{Moser2021} showed different assumptions made about the sample could affect how the resulting profile is interpreted, and in the pipelines for analyzing the observed profiles (i.e., inferring astrophysical parameters by fitting to a model). When forward modeling it is important to have the most accurate interpretation of the observations possible, as understanding a feature of the data is crucial to knowing whether it is due to the experiment configuration or is actually relaying astrophysical information. For example, the tSZ signal has a nonlinear dependence on mass (proportional to $M^{5/3}$), so it is important to match the observed mass distribution of the sample in the modeling rather than model quantities for a single value, e.g., the average mass of the sample (see \citet{Moser2021} for more details). Other examples include correctly modeling the experiment's beam profiles and response functions and estimating two-halo term contributions.

There is a pipeline previously developed to accomplish the forward modeling of observed SZ profiles, called \texttt{Mop-c-GT}\footnote{\url{https://github.com/samodeo/Mop-c-GT}}, introduced to analyze the recent Atacama Cosmology Telescope (ACT) data (Data Release 5) presented in \citet{Amodeo2021}. As a brief overview, this repository inputs a spherically-symmetric model for gas density and pressure (either computed from a model or calculated from a simulation), projects the profile along the LOS, and convolves the projected profile with any instrumental beam to produce modeled observed SZ profiles. In \citet{Amodeo2021}, this pipeline was used as a likelihood to fit the observed SZ profiles for astrophysical parameters of two models: the generalized Navarro-Frenk-White model (GNFW, \citet{gnfw}) and the Ostriker-Bode-Babul model (OBB, \citet{Ostriker2005}). 

Additionally, \texttt{Mop-c-GT} was used to forward model stacked SZ profiles for halos selected from the IllustrisTNG (TNG) simulation \citep{Marinacci2018,Naiman2018,Pillepich2018,Springel2018,Nelson2018a,Nelson2019} and compared to the ACT profiles. Described further in \citet{Moser2021}, the halo sample from the simulation was selected to most closely match the known properties of the modeled observed sample in terms of mass range and weighted according to the observed mass distribution. In this case, the modeled sample was the CMASS (``constant- mass") galaxy sample of the Baryon Oscillation Spectroscopic Survey (BOSS), Data Release 10 \citep{Ahn2014} with a stellar mass range of $10.7 \lesssim \log{(M^*/M_{\odot})} \lesssim 11.7$ \citep{Maraston2013}. It was shown that the simulated halos could reproduce the observed kSZ profile well (see Figure 7 of \citet{Amodeo2021}), but the simulated tSZ profiles were significantly different from the observed tSZ profile.

Along with the general goal of improvement of methods for interpreting future observations, the discrepancy between the simulated tSZ profiles and ACT data discussed in \citet{Amodeo2021} serves as the motivation of this study. There are many uncertainties involved in this process, in both simulations and interpreting observations, so here we explore a number of possible systematics on both ends to see which could potentially explain a discrepancy such as this one. On the side of uncertainties in interpreting observations, we test (and update) calculations within the \texttt{Mop-c-GT} pipeline in Section~\ref{sec:projection_effects}. This includes various computations such as varying the LOS integration distance and beam convolution methods. On the side of uncertainties within the simulations, we test the effects of an uncertain halo selection mass range in Section~\ref{sec:mass_uncertainty}. Here we select halos from the TNG simulation varying the limits of the mass ranges modeled in \citet{Amodeo2021} up to $20\%$ to see if the resulting shift in profiles could possibly account for the difference in observations. Finally, in Section~\ref{sec:compare_to_obs} we compute updated forward modeled profiles with all of the corrections and possible changes to the pipeline discussed in prior sections and plot them against the ACT profiles of \citet{Amodeo2021} that first showed this difference.

\section{Projection Effects}\label{sec:projection_effects}
Observed kSZ and tSZ signals are integrated along the LOS and include beam effects from the instrument, so to accurately forward model profiles, we need to project the profile and convolve the beam similarly. The repository called \texttt{Mop-c-GT} was developed to forward model SZ profiles, and was recently used in \citet{Amodeo2021} and \citet{Schaan2021} to analyze SZ data from ACT DR5. This pipeline inputs a model for a spherically-symmetric stacked gas density or pressure profile, projects along the LOS, and convolves the projected profile with any calculated or forecasted experiment beam. It also has the functionality to apply aperture photometry filters to remove any potential foreground effects. There are several calculations involved in this process, so in this section, we explore various potential systematics and their effects on the resulting modeled SZ profiles. 

As shown in \citet{Moser2021}, it is important to include any contributions to the signal from neighboring halos, known as the two-halo term, when forward modeling SZ profiles. In this study, all of the following tests compute example profiles as the addition of the signal from the main halo, known as the one-halo term, and the two-halo term. We use a generalized Navarro-Frenk-White (GNFW, \citet{gnfw}) model for the one-halo term, with parameters for an average halo mass of $10^{13} M_\odot$ and redshift $z=0.55$ (see \citet{Battaglia2012a} and \citet{Battaglia2016} for the equations used to calculate GNFW parameters as functions of mass and redshift). We use a two-halo term model from \citet{Vikram2017}, and add it to the one-halo term to calculate the total signal, shown in Equation~\ref{eq:gnfw_2h}. 

\begin{equation}
    \label{eq:gnfw_2h}
    \begin{aligned}
    \rho_{GNFW} &= \rho_{1h} + A_{k2h}\rho_{2h} \, , \\
    P_{GNFW} &= P_{1h} + A_{t2h}P_{2h}
    \end{aligned}
\end{equation}
where $\rho_{GNFW}$ and $P_{GNFW}$ are the total modeled profiles for density and pressure, respectively, $\rho_{1h}$ and $P_{1h}$ are the one-halo term contributions, $A_{k2h}$ and $A_{t2h}$ are the amplitudes of the two-halo term we set equal to 1.0 in this study, and $\rho_{2h}$ and $P_{2h}$ are the two-halo term contributions. 

\begin{figure*}
    \centering
    \includegraphics[scale=0.65]{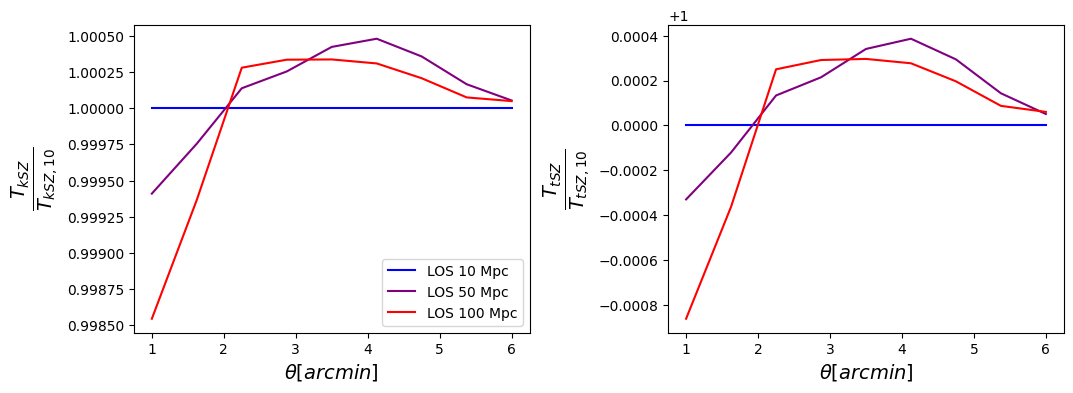}
    \caption{Ratios of example profiles (kSZ left, tSZ right) calculated with GNFW model parameters \citep{Battaglia2012a,Battaglia2016} for mass $10^{13} M_\odot$ and $z=0.55$ integrated out to different LOS distances. The default integration distance used in \citet{Amodeo2021} was LOS = 10 Mpc, so we normalize the other profiles to see the effects with respect to this default profile. It can be seen that changing this distance in the computation of the projection does not have a significant effect on the resulting profile, on the order of $\ll 1\%$.}
    \label{fig:LOS}
\end{figure*}

\subsection{LOS Expansion}
First, \texttt{Mop-c-GT} cylindrically projects the gas density or pressure profile along the LOS. To create or interpret profiles, previous studies \citep{Amodeo2021,Schaan2021,Moser2021} used stacked profiles from multiple hundreds of halos. These halos could have differing individual orientations, but by stacking a large number of these objects we are able to assume a spherically-symmetric density and pressure profiles \citep[e.g.,][]{Battaglia2012a}. Thus, changing the direction of projection should not yield different results. However, there could possibly be effects from increasing the LOS distance along which we integrate to project the profiles. This follows the idea that there could perhaps be additional, halo-correlated contributions along the LOS that would be detected along the larger integration axis. To test this possibility, we explore how changing the LOS of integration affects the resulting profiles. 

In previous studies, the profiles were integrated along a LOS within the range $10^{-3} - 10$ Mpc in front of and behind the halos. In this study, we compute example profiles with the integration axis extended out to 50 Mpc and 100 Mpc. The results are shown in Figure~\ref{fig:LOS}, with ratios of the profiles (kSZ left, tSZ right) to the profile computed with the default LOS distance of 10 Mpc. It can be seen from this figure that the effects of this change in LOS distance are completely negligible, $\ll 1\%$, which highlights the benefit of using the aperture photometry filter. 
We emphasize that the projection involves the sum of one-halo and two-halo terms (see Equation~\ref{eq:gnfw_2h}) while random components along the LOS are filtered out by aperture photometry.

As we increase the LOS distance, we can see more of a decrement in the inner profile, which could possibly be due to the aperture photometry removing more of the signal, and a slight increment in the central to outer profile. These extremely small differences are not able to account for the significant discrepancy between simulated and observed profiles, and support the robustness of the result discussed in \citet{Amodeo2021}.

\subsection{Beams}
After projecting along the LOS, the profile is convolved with any instrumental beam, either forecasted or estimated from a real experiment, to show how a given instrument or experiment would observe the profile. In previous studies, the profiles were convolved with the same beams with which the ACT+Planck coadded maps are convolved. As described in \citet{Naess2020}, these beams have non-Gaussian and scale-dependent profiles. They were estimated through a system of equations that inputs maps of different noise properties and beams from the data that make up the ACT+Planck coadded maps, then solves for the one beam that satisfies this system. Following this process, \citet{Naess2020} estimated that these beams were accurate to 10\%.

We have estimated the beams an independent way by stacking on point source signals in the ACT+Planck coadded maps. This is a data-driven method to determine the average beam in each of the ACT+Planck coadded maps. The sources are bright ACT point sources and provide measurements of the beam profile on small angular scales. On large scales the measurement of the beam profile from sources becomes noisier, since the signal decreases. However, on large scales the signal and the beam in the ACT+Planck coadded maps are dominated by the Planck data \citep{Naess2020} and the Planck beam is measured very precisely \citep{Planck2018}. Thus, we normalize our measured beam to match the Planck beam on large scales.

%DR5 beams are the target beam that I ask the equation to produce. [In more detail] It's an equation system that takes in maps of different noise properties and beams, and produces an output map with the beam one asks for
 
%This shows the importance of an accurate estimation of the beam profile, as it results in significant differences in any modeled profile that is calculated with beam convolution. 

\begin{figure*}[t]
    \centering
    \includegraphics[scale=0.65]{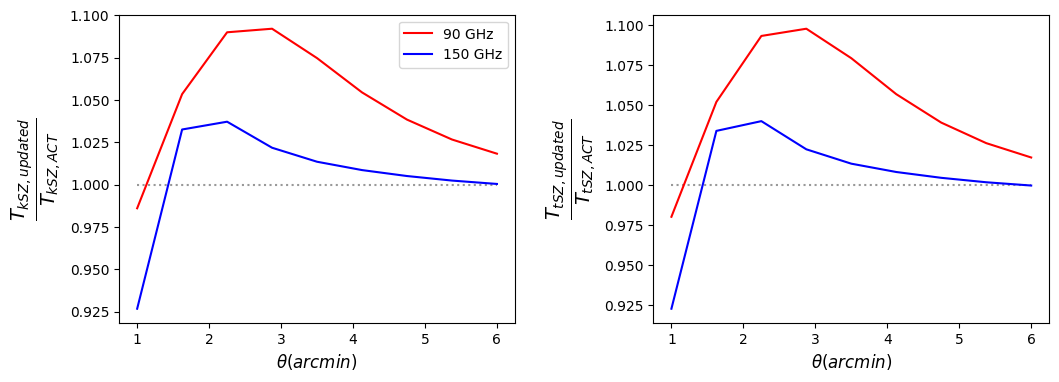}
    \caption{Ratios of kSZ (left) and tSZ (right) profiles convolved with the updated beam to the profile convolved with the ACT beam used in \citet{Amodeo2021}. The red curves show profiles calculated at 90 GHz, and the blue curves show profiles calculated at 150 GHz. Profiles computed at both frequencies show differences $\sim10\%$ using the updated beam, so this is an important systematic to consider when modeling profiles of a given experiment.}
    \label{fig:beams_act_updated}
\end{figure*}

In Figure~\ref{fig:beams_act_updated} we show the differences in profiles resulting from convolution with the beams used in \citet{Schaan2021} and \citet{Amodeo2021} and the updated beam. 
%The projection process for the profiles convolved with the ACT beam is FFT (see Section~\ref{sec:beam_implementation}). 
At both frequencies of 150 GHz and 90 GHz, convolution with the updated beam can have an effect on the profiles up to $\sim10\%$, showing that having an accurate estimation of the beam is important. 
%This is an important systematic to be considered in future studies, but changes of this magnitude are not significant enough to change the results discussed in \citet{Amodeo2021}.

\subsection{Beam Implementation}\label{sec:beam_implementation}
The method with which we convolve the profiles with the beam should not be important, but as another potential source of inaccuracy, we explore two different convolution methods. The method used in \citet{Amodeo2021} was the Fast Fourier transform (FFT), and here we add the Fast Hankel transform (FHT) method to compare. In theory, these methods should yield similar profiles, but the FHT should be faster and potentially more accurate than FHT, for reasons described further below.

\subsubsection{FFT Beam Convolution Method}\label{sec:FFT_transform}
The method used in \citet{Amodeo2021} is the FFT, which constructs a grid of points specified by size in arcminutes and number of pixels over which the profiles are convolved. This is potentially not as accurate as either performing the integrals analytically or using the FHT method because we have to estimate the size and parameters of this map, leading to potential discrepancies in the profiles. In \citet{Amodeo2021}, the FFT map size was set to $15^\prime$ and the number of pixels was set to 500 on each side, so here we test the effects of varying these parameters on the resulting profiles.

\begin{figure*}
    \centering
    \includegraphics[scale=0.65]{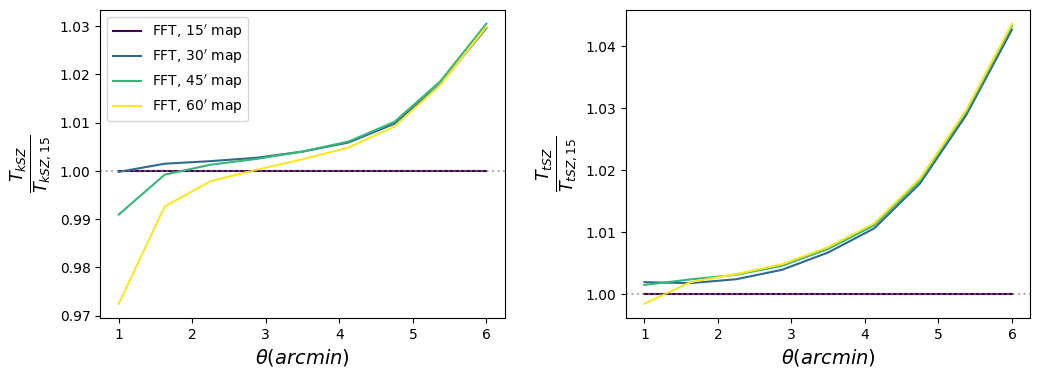}

    \includegraphics[scale=0.65]{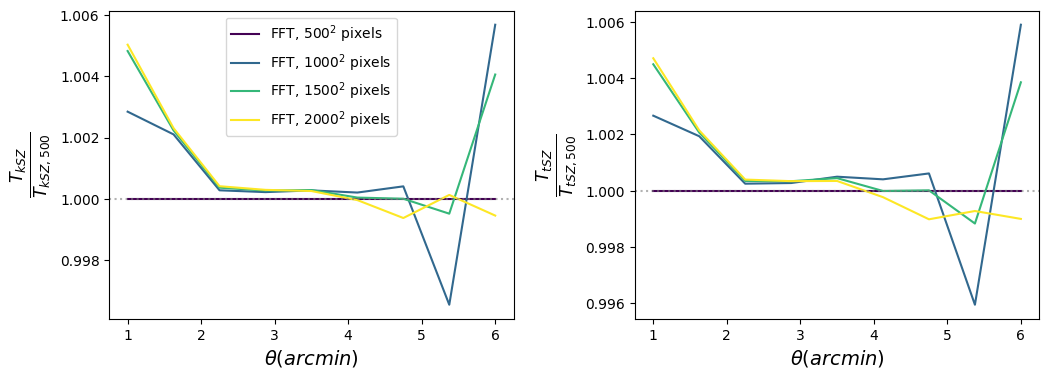}
    \caption{This figure shows the two systematics of the FFT beam convolution method discussed in Section~\ref{sec:FFT_transform}. The top line shows ratios of kSZ (left) and tSZ (right) profiles varying the size of the FFT map with respect to the default value used in \citet{Amodeo2021} ($15^\prime$). We see differences on the order of $\sim5\%$ in the outer profile for the different map sizes. The bottom line show the ratios of the profiles computed with map size $15^\prime$ but varying resolutions. We see even smaller differences, sub $1\%$, varying the resolution of the FFT map used to convolve the profiles. These figures show that these systematics, while important to implement correctly, do not have significant effects on the profiles.}
    \label{fig:fft_map_size_resolution}
\end{figure*}

First, we vary the size of the grid while keeping the resolution fixed as the same used in \citet{Amodeo2021} $(15^\prime-500^2)$. The top line of Figure~\ref{fig:fft_map_size_resolution} shows the differences in the profiles, with ratios of kSZ (left) and tSZ (right) profiles to the profile computed with the default value, map size of $15^\prime$. It can be seen that the larger maps seemingly converge to a larger value in the outer profile, indicating that using this FFT method and a map that does not extend far enough can result in a loss of signal on the order of $\sim5\%$. This is a relatively small effect, so does not account for the differences we see in the simulated tSZ profiles compared to the observations. However, as the precision in observations continues to improve, these small order effects will become important to correctly model. 

Next, we test the importance of resolution when constructing the grid of points for the FFT convolution. In \citet{Amodeo2021}, with a map size of $15^\prime$, the number of pixels was set to be $500^2$. In the top line of Figure~\ref{fig:fft_map_size_resolution} we conserve the resolution by increasing the number of pixels and map size by the same factor, i.e. when doubling the map size from $15^\prime$ to $30^\prime$, we increase the number of pixels on a side from 500 to 1000. In the bottom line of Figure~\ref{fig:fft_map_size_resolution} we leave the map size set at $15^\prime$ but vary the number of pixels. It can be seen that the resolution of the FFT grid does not have a significant effect on the profiles, with the largest differences in the outer radial bins having magnitudes sub $1\%$.

With the goal of trying to better match the simulated profiles to the observed profiles, when using the FFT method in future computations we will use a larger map size of $30^\prime$ with $1000^2$ pixels.

\subsubsection{FHT Beam Convolution Method}\label{sec:FHT_transform}
The FHT transforms between radially symmetric profiles in real space and Fourier space. It is more useful than the FFT in certain physical situations that span several orders of magnitude, such as extended profiles, due to its use of logarithmically-spaced input points rather than linearly-spaced \citep{Talman1978,Hamilton2000}. The FHT does not rely on an input grid of points like the FFT method, so we do not have to worry about inaccuracies due to the systematics in defining the map as discussed above in Section~\ref{sec:FFT_transform}. Additionally, without an input grid of points and computation time scaling of $Nlog(N)$ (where $N$ is the number of points in the convolution), it should be much faster than the FFT method. For single calculations this is not significant, but when running Markov chain Monte Carlo calculations, the saved computation time will make analyzing SZ profiles much more efficient.

In this study, we use a version of the FHT implemented in the package \texttt{pixell}\footnote{\url{https://github.com/simonsobs/pixell}}.
\texttt{Pixell} is used for analyzing maps of the sky, including CMB intensity maps, and has functionality to perform different kinds of transforms. A method to compute the FHT was recently added using scipy's fast Hankel transform. This method does not need to input the radial values of the profile like the FFT method, but rather inputs a range of harmonic space $\ell$ values over which the profile needs to be calculated. This is the main systematic we will be testing for this method, to see how varying both limits of the $\ell$ value range affects the resulting profile.

For starting points, we use the real-space limits of our SZ profiles to estimate what the limits of the $\ell$ range should be. Using the relations $R_{rad} = \frac{R_{\rm Mpc}}{AngDist(z)}$, where $AngDist(z)$ is the angular distance as a function of redshift and $R_{rad} = \frac{1}{\ell}$. We estimate $R_{rad}$ for $10^{-3} \leq R_{\rm Mpc} \leq 10$, and we obtain an $\ell$ range of $[135, 1.4e6]$ at redshift $z = 0.55$. We use these values as our starting points for the following tests.

\begin{figure*}
    \centering
    \includegraphics[scale=0.65]{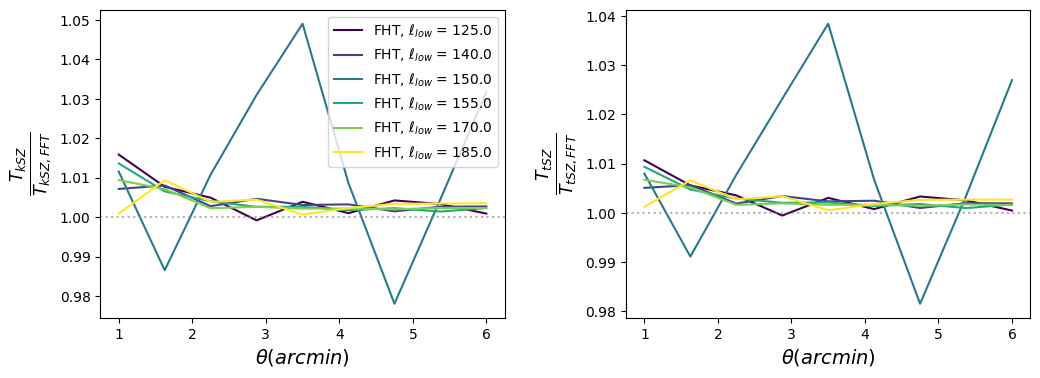}

    \includegraphics[scale=0.65]{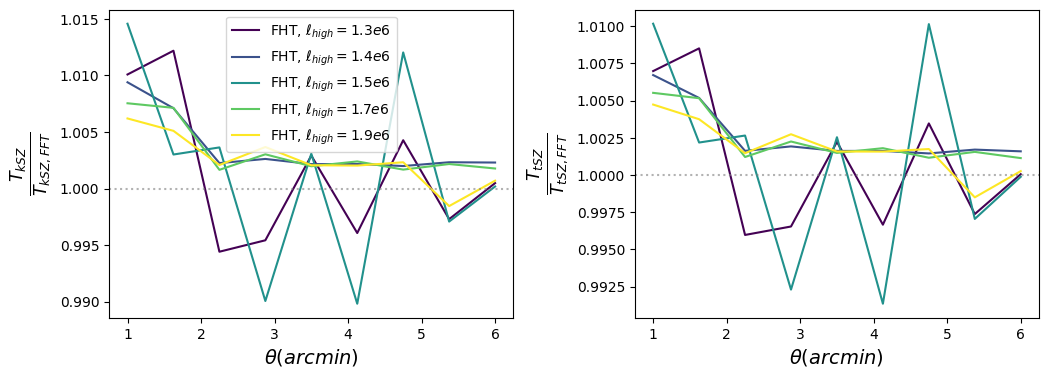}
    \caption{This figure shows the systematics of the FHT beam convolution method discussed in Section~\ref{sec:FHT_transform}. The top line shows ratios of kSZ (left) and tSZ (right) profiles varying the lower limit of the $\ell$ range input into the FHT calculation with respect to the profile computed with the FFT method (map size $30^\prime$, pixels $1000^2$). The FHT profiles in this line are calculated with an upper $\ell$ limit of $1.4e6$. The bottom line shows the same, but varying the upper limit of the $\ell$ range. Here, we leave the lower limit fixed at 170 and vary around the $1.4e6$ value estimated by the real-space limit. There are, of course, many other values that could be shown for both of these tests, but the main result is that once some exploration of this nature has been done to narrow down both limits of the ranges, the specific value does not have a significant effect on the profiles.}
    \label{fig:hankel_vary_lrange}
\end{figure*}

First, we vary the lower limit of the $\ell$ range going into the FHT while keeping the upper limit fixed at $1.4e6$, shown in the top line of Figure~\ref{fig:hankel_vary_lrange}. We compute these example GNFW SZ profiles using the FHT method, and take the ratio of the example profiles to the profile computed through FFT (with map size $30^\prime$, number of pixels $1000^2$). Based on the real-space limits, we estimate the value of $\ell_{low}$ to be around 135, but we vary the lower limit in increments of 5 covering the range $125 \leq \ell_{low} \leq 200$ to test a wider range. Most of these values result in good matches to the profile computed with the FFT method, with differences on the order of $\sim1\%$. However, some of these values result in profiles with differences on the order of $\sim5\%$, so it is important to test further before settling on a value.

Next, we vary the upper limit of the $\ell$ range going into the FHT. Based on the previous test, we select $\ell_{low} = 170$ as the lower limit, as it was one of the values resulting in the best match to the profile computed through the FFT method. The estimated value for the upper $\ell$ value we calculated previously using the real-space limits was $1.4e6$, so we test a few other values around this and plot the best matches to FFT in Figure~\ref{fig:hankel_vary_lrange}. Similarly to the test varying $\ell_{low}$, there are values not shown here that are worse matches, so it is important to explore the range before adopting a value. However, there are several values that result in profiles of nearly equally good matches to FFT, so choosing one specifically out of these is not that important. Therefore, moving forward we will use $\ell$ range $170 \leq \ell \leq 1.4e6$ when calculating profiles with the FHT method.

\subsection{Radial Limits}\label{sec:radial_limits}
Rather than vary certain parameters set in \texttt{Mop-c-GT} for the calculations in \citet{Amodeo2021}, in this section we test for inaccuracies in the original pipeline and suggest replacements for future computations.

\subsubsection{Radius Definition Correction}\label{sec:radius_definition}
A possible systematic in modeling the SZ profiles is to have mismatched radii at different computational steps along the pipeline. It is very important to select the radii of interest correctly, along with the ranges of any interpolated radii. We perform tests to ensure all radial limits within the profile calculation are aligned. In the \citet{Amodeo2021} and \citet{Schaan2021} studies, the method for defining the profiles' radii was to set the radial extent based on the observed radial bin $\theta$ (the x-axis in all of the prior figures). Then, to perform the beam and response function convolutions using the FFT method, \texttt{Mop-c-GT} defined and interpolated a map with which the beam was convolved. As previously described, this map may not have extended out far enough (see Figure~\ref{fig:fft_map_size_resolution}), resulting in slight discrepancies in the computed profiles.  

\begin{figure*}
    \centering
    \includegraphics[scale=0.65]{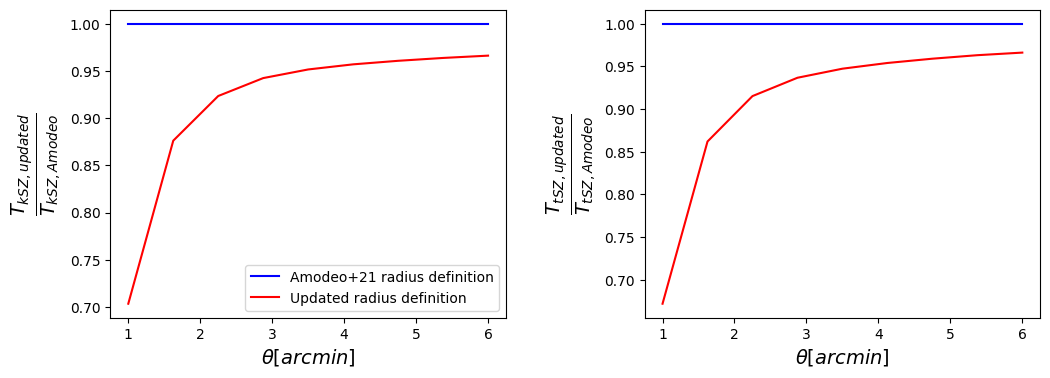}

    \includegraphics[scale=0.65]{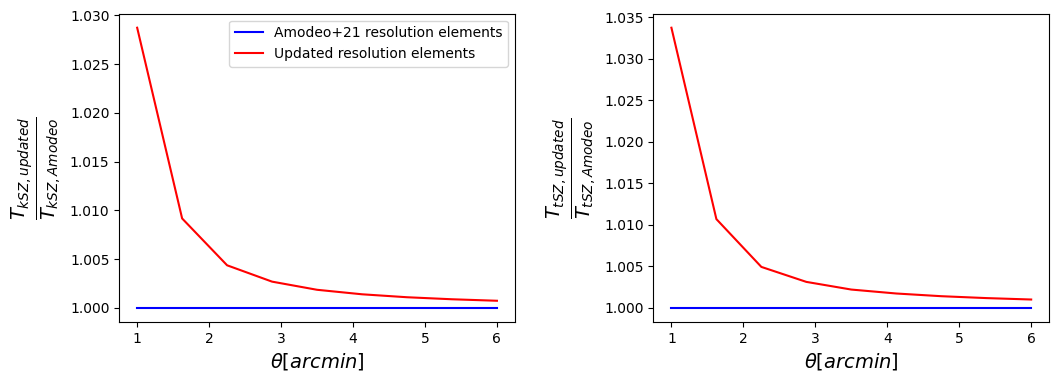}
    \caption{This figure shows the importance of correctly matching the radial limits along each step of the pipeline, discussed in Section~\ref{sec:radial_limits}. The top row shows the correction for the radius definition described in Section~\ref{sec:radius_definition}, with ratios of profiles computed using the updated radius definition (red) to the method used in \citet{Amodeo2021} (blue). It can be seen that having the convolution radius extending out beyond the input radius results in loss of signal and amplitude decreases on the order of $\sim30\%$. The bottom row shows the correction for the resolution elements being used described in Section~\ref{sec:resolution_elements}, with ratios of profiles computed using the updated method (red) to the method used in \citet{Amodeo2021} (blue). It can be seen that this correction is not as significant as the radius definition correction, but does result in profile differences on the order of $\sim3\%$.}
    \label{fig:radial_limits}
\end{figure*}

We test a new method for defining radius arrays that will not miss any profile information by calculating the profiles on the entire radial range defined in the FFT map rather than only going out to the observed $\theta$ bin. This means that regardless of the radial bin $\theta$ at which we want to estimate the SZ profiles (in the range 1'-6', shown by the x-axis of all the prior figures), we first compute the profile LOS projection up to 30', then we take the value corresponding to the radial bin of interest, and finally we convolve the projected profile with the beam, using a FFT map of size 30'. For this test, we compute profiles using both radius definition methods, shown in the top line of Figure~\ref{fig:radial_limits}. The red lines show the ratio of the profile computed with the updated method to the profile computed with the method used in \citet{Amodeo2021} (shown in blue). It can be seen that the updated method results in a decrease of amplitude of the profiles, especially in the inner radial bins. 
%\textbf{I feel like the intuitive thing to think would be if we are missing part of the profile using the old way, shouldn't our corrected profile be higher in amplitude? In reality, the fact that we were missing some of the profile before resulted in flat interpolations at values HIGHER than they should have been, thus we were seeing higher values then, and lower values using the corrected method now.}
This is due to the method used in \citet{Amodeo2021} not extending out far enough, so when the beam convolution was interpolating over radii beyond what was given by the data, it returned a flat line at a profile value higher than it should be. The new method preserves the shape of the profile out to larger radii, which continues to decrease as radius increases rather than going flat, thus we see lower profile values using the new method.
%that needs to be formalized better, but fine for now

In relation to the problem discussed in \citet{Amodeo2021} of the simulations under-predicting the data, this correction actually makes the comparison worse, as it moves the profiles down in amplitude when a higher tSZ amplitude would be needed to match the observations. However, this is only looking at one of the corrections independently; when combined with all of the other corrections presented in this study, it does not have a significant effect on the profile (see Figure~\ref{fig:obs_compare}). 

\subsubsection{Resolution Elements Correction}\label{sec:resolution_elements}
In testing the different radial limits, we found another potential improvement to the projection process used in \citet{Amodeo2021}. Certain arrays in the pipeline are defined to have a given number of elements spanning the observed $\theta$ range. In subsequent array definitions, the number of elements is increased to have narrower $\theta$ bins, thus increasing the resolution when performing the beam convolution. The way these arrays were defined led to the resolution enhancement extending the integration limit farther than intended.

Therefore, we implement an updated method that leaves the resolution enhancement, but corrects the unintended effect of changing the radial limits for the integration. We test the impact of this updated method on the profiles, shown in the bottom row of Figure~\ref{fig:radial_limits}. Similarly to the top line, the red line shows the ratio of the profile computed using the updated method to the profile computed using the method from \citet{Amodeo2021} shown as the blue line. It can be seen that this correction leads to small differences in the inner profile, of the order of $3\%$. Similarly to other corrections in this study, this is not a large correction but as our observations achieve higher sensitivities, these kinds of small-order modeling effects will become increasingly important to implement correctly.

\section{Mass Selection Uncertainties}\label{sec:mass_uncertainty}
As discussed in \citet{Moser2021}, modeling halo samples for the purpose of constraining baryonic feedback processes has many uncertainties that are important to consider. Previous works \citep{Amodeo2021,Moser2021,Moser2022} constructed halo samples from the IllustrisTNG simulation to have properties that most closely match those of the observed sample being modeled. In this case, we have been modeling the CMASS sample from BOSS, and have attempted to model the observed sample's properties of mass ranges and distributions. 

There are multiple systematics when trying to select for mass ranges of the simulations. First, it is well-known that the stellar-halo mass relation (SHMR) is uncertain, and the shape of this relation differs among different samples of galaxies, observed or simulated (e.g., \citet{Shankar2014}). Halo masses for the CMASS sample were calculated from the stellar masses of \citet{Maraston2013} using the \citet{Kravtsov2018} SHMR relation, but \citet{Moser2021} showed that TNG's SHMR has a very different shape. This leads to uncertainty in selecting a mass range from the simulations described by a different SHMR to try to best model the observed sample. Therefore, there are systematics in both selecting a mass range from the simulations and in determining masses of the observations based on the assumed shape of the SHMR being imperfect. For these reasons, we explore the effects of mass uncertainty on the profiles by varying the masses being selected from the simulations to account for systematics on either end. 
  
\begin{figure*}
    \centering
    \includegraphics[scale=0.65]{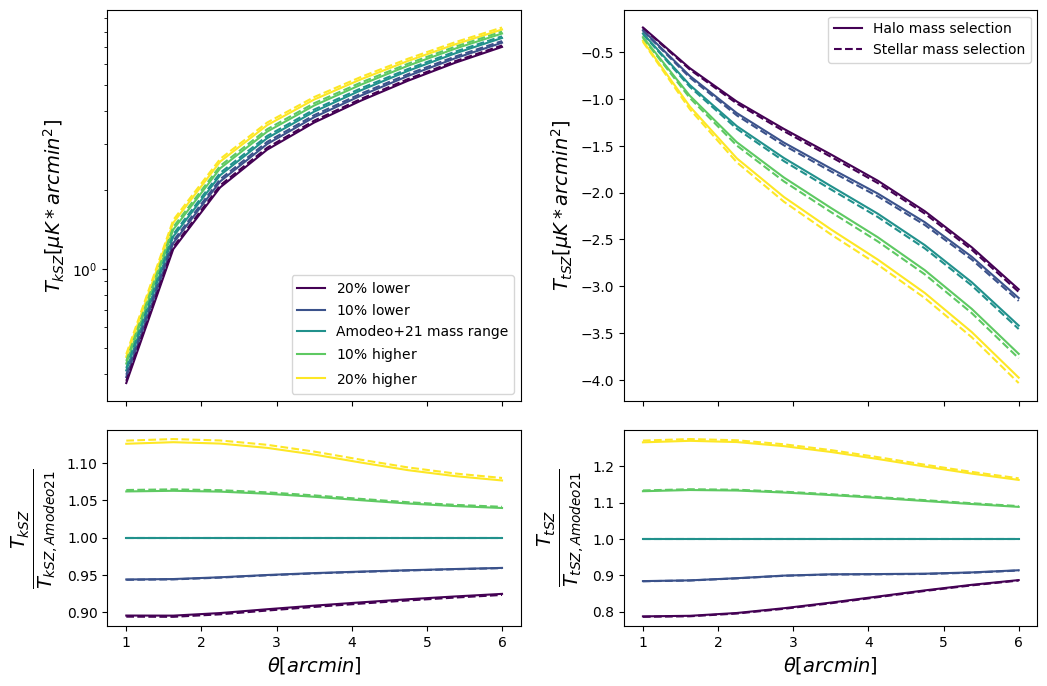}
    \caption{This figure shows the mass selection uncertainties of selecting simulated halos to model an observed sample. These SZ profiles were all calculated from the TNG simulation for different mass ranges in relation to the mass selection range used in \citet{Amodeo2021}. We decrease and increase the mass range limits up to $20\%$, and show the differences between the selections of different mass type (solid for halo mass, dashed for stellar mass) within the simulation as well. It can be seen that the uncertainty in mass can lead to significant differences in the resulting profiles, roughly equivalent to the percentage difference in the mass limits.}
    \label{fig:mass_ranges}
\end{figure*}

In Figure~\ref{fig:mass_ranges} we show forward-modeled profiles calculated from halos of the TNG simulation varying the limits of the mass range chosen for the profiles in \citet{Amodeo2021} up to $20\%$. More specifically, we choose a stellar mass range of $9.5 \lesssim \log{(M^*/M_\odot)} \lesssim 11.7$ and halo mass range of $11.3 \lesssim \log{(M_h/M_\odot)} \lesssim 13.9$ and vary both limits up to $20\%$. We note that the studies of \citet{Moser2021} and \citet{Moser2022} further limit the mass range to not include the lower mass objects with $\log{M^*/M_\odot} \leq 10.7$, but the results are largely the same, since the lower mass objects are significantly down-weighted in the mass-weighting method used in \citet{Amodeo2021}.
%\emm{Does it matter that these are different from the Maraston2013 histogram that I quote above? The histogram shows $\sim10.5-12$ for mstar which is why above I say 10.7-11.7, but we used different ranges for Amodeo21. In Moser21 and Moser22 we use 10.72-11.72 for ms and 12.12-13.98 for mh.} 
We also show profiles using different mass type selections from the simulation, with the solid lines showing the halo mass-selected sample and the dashed lines showing the stellar mass-selected sample. It can be seen that this uncertainty in mass range can have a significant effect on the resulting modeled SZ profiles, particularly for tSZ. For kSZ we see around $10\%$ differences at the largest, while for tSZ we see up to $20\%$ differences for the shifted mass ranges. Clearly this is an important systematic to keep in mind while trying to forward model SZ profiles, however the differences are still not significant enough to account for the discrepancy for the simulated tSZ profiles with the observations (see Figure~\ref{fig:obs_compare}) and support the result found in \citet{Amodeo2021}.

It is important to note that while the changes in amplitude of the profiles resulting from this systematic are significant, we cannot simply continue to increase the amplitude of the tSZ profile in hopes of better matching the observed tSZ data without also increasing the amplitude of the modeled kSZ profile. Theoretically, if we were to change the tSZ model somehow to align with the tSZ data, our kSZ model would no longer yield as close of a match to the kSZ data that we currently calculate. Therefore, something in the tSZ model or simulations used in computing modeled tSZ profiles needs to be changed without affecting the fit for the modeled kSZ profiles.

\section{Comparison to Observations}\label{sec:compare_to_obs}

\begin{figure*}
    \centering
    \includegraphics[scale=0.65]{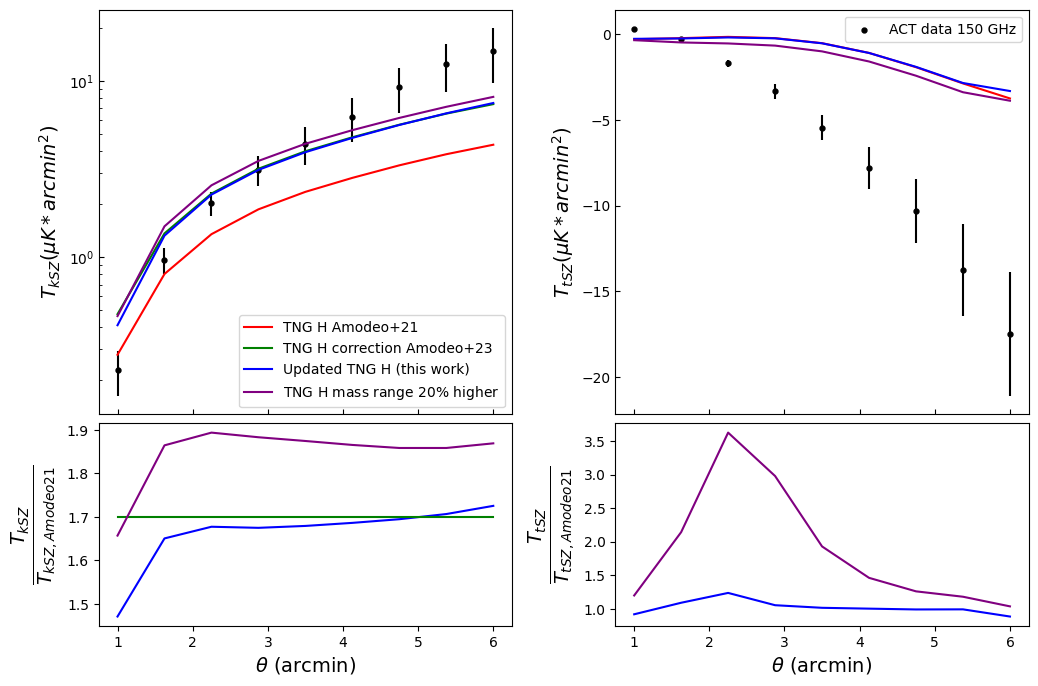}
    \caption{This figure is an updated version of Figure 7 in \citet{Amodeo2021} showing the comparison of simulated kSZ (left) and tSZ (right) profiles to ACT data. The black points and errorbars show the ACT data at a frequency of 150 GHz, the red lines show the original forward-modeled TNG profiles shown in \citet{Amodeo2021}, the green line for kSZ shows the original TNG profile corrected with the missing factor discussed in \citet{Amodeo21erratum}, and the blue lines show the updated TNG profiles with all of the changes to the pipeline described in this study. These updated curves were computed with a LOS distance of 10 Mpc, beam convolved through FFT method with map size $30^\prime$ $1000^2$ pixels, updated ACT beam profile, and corrected radius definition and resolution elements methods. Finally, the purple lines are TNG profiles allowing $20\%$ higher mass limits in halo mass selection to show that even with this magnitude of uncertainty, none of these profiles are able to match the tSZ observations.}
    \label{fig:obs_compare}
\end{figure*}

The main motivation of this study is to explore whether the discrepancy between simulated SZ profiles and ACT observations presented in \citet{Amodeo2021} is due to systematics along the forward modeling process. %or something missing in either the simulations or observations. 
We have tested several computations and details along the forward modeling pipeline \texttt{Mop-c-GT} used in this previous study. We have varied parameters and corrected methods and found varying degrees of importance on the resulting profiles. Now we combine all of our findings to calculate `improved' profiles from TNG with the following properties: LOS distance of 10 Mpc, convolved with updated ACT beams through FFT with map size $30^\prime$ and $1000^2$ pixels, corrected radius definition, corrected resolution elements, and corrected the missing factor of 1.7 for kSZ profiles discussed in \citet{Amodeo21erratum}.

We show the comparison of the updated profiles to the profiles calculated in \citet{Amodeo2021}, and show the comparison of both to the ACT observations in Figure~\ref{fig:obs_compare}. The top line shows the kSZ profiles (left) and tSZ profiles (right), and the bottom panels show the ratio of the different profiles to the original TNG profile presented in \citet{Amodeo2021}. The black data points and error bars show the ACT data at a frequency of 150 GHz, the red line shows the original TNG line presented in \citet{Amodeo2021} (here we only show the halo mass-selected sample, as Figure~\ref{fig:mass_ranges} showed the mass type selection yielded similar results), the green line for kSZ shows the original TNG line corrected with the missing factor discussed in \citet{Amodeo21erratum}, and the blue line shows updated TNG profiles with all of the changes listed above. Finally, we plot the yellow lines of Figure~\ref{fig:mass_ranges} expanding the mass selection range from TNG to be $20\%$ higher in purple to show that even the profiles with up to $20\%$ higher amplitudes allowed by the uncertainty in mass still cannot account for the discrepancies we see between simulations and observations. 

\section{Conclusions}
Complete forward modeling of SZ profiles is an involved process with many systematics to consider. An important result of \citet{Amodeo2021} was the comparison of forward-modeled SZ profiles computed by stacking halos of the IllustrisTNG simulation to ACT data presented in \citet{Amodeo2021} and \citet{Schaan2021}. This comparison showed a large discrepancy between the simulated tSZ profiles and the data, so in this study we further tested several of the methods along the forward modeling pipeline. These tests include the effects of integration LOS distance, in which increasing the LOS from 10 Mpc to 50 Mpc and 100 Mpc led to $\ll1\%$ changes in the profile, see Figure~\ref{fig:LOS}. We tested the importance of estimating an accurate beam profile, in which we convolved the example profiles with the ACT beam used in \citet{Amodeo2021} and with an updated beam estimated using a new method, and found changes in the profiles
on the order of $10\%$, see Figure~\ref{fig:beams_act_updated}. We tested the beam convolution method, in which we convolved the profiles with the beam using FFT and FHT, and found profile changes on the order of $5\%$, see Figures~\ref{fig:fft_map_size_resolution} and~\ref{fig:hankel_vary_lrange}. We corrected various definitions of radii and resolution elements along the pipeline to ensure that all of our radii were calculated correctly, and found profile changes on the order of $30\%$, see Figure~\ref{fig:radial_limits}. Lastly, we explored the uncertainty in mass selection to see if perhaps the mass limits used to select the TNG halos in \citet{Amodeo2021} differed up to $20\%$ in either direction the profiles would be shifted enough to account for the differences in profiles. This resulted in profile changes on the order of $20\%$, see Figure~\ref{fig:mass_ranges}. All of these corrections (aside from the uncertainty in mass range) have been integrated into the upcoming Simons Observatory (SO) likelihood repository, \texttt{SOLikeT}\footnote{\url{https://github.com/simonsobs/SOLikeT}}. This repository will become public soon, and will contain several likelihoods and theory calculations for future SO data analysis.

In Figure~\ref{fig:obs_compare} we combined the corrections in all of these methods to calculate an updated TNG profile, allowing for uncertainty in the mass selection, and showed that our updated profiles still could not reconcile the differences with the observations. This indicates that the original results of \citet{Amodeo2021} were robust, and not due to the systematics in profile calculation we have tested here. There are other methods that could be tested, such as further explorations of the two-halo term present in the simulations and in the model we use from \citet{Vikram2017}. We could also perform the same kind of analysis at different redshifts (comparing to other observed samples) to see if there is any trend with redshift. This is not an exhaustive list of further tests that could be performed, and we leave any additional possible tests for future studies.

As the signal-to-noise in measurements continues to increase, correctly modeling these kinds of details of the sample and in calculating the profiles will become necessary.
Moving forward it is important to better understand this problem, as the observations still pose a challenge for the numerical methods used to describe the CGM/IGM heating in simulations.

\section*{Acknowledgements}
We thank Gordon Stacey and Sigurd Naess for their helpful comments. E.M. and N.B. are supported by NSF grant AST-1910021 and NB acknowledges support from NASA grants 80NSSC18K0695, 80NSSC22K0410, and 80NSSC22K0721.

\bibliographystyle{aasjournal}
\bibliography{cits.bib}
\end{document}